\documentstyle[11pt,newpasp,twoside,psfig]{article}
\begin{document}

\title{Reverberation Mapping of High-Redshift, High-Luminosity Quasars}
\author{Shai Kaspi, Hagai Netzer, Dan Maoz and Ohad Shemmer}
\affil{School of Physics and Astronomy, Tel-Aviv University, Tel-Aviv, Israel}
\author{W. N. Brandt and Donald P. Schneider}
\affil{Department of Astronomy and Astrophysics, The Pennsylvania State University, University Park, PA 16802, USA}

\setcounter{page}{1}
% this page number will be filled later by the editors....
\index{Kaspi, S.}
\index{Netzer, H.}
\index{Maoz, D.}
\index{Brandt, W. N.}
\index{Schneider, D. P.}

\begin{abstract}
We present preliminary results from a reverberation mapping program to
measure the Broad Line Region size in high-redshift, high-luminosity
quasars. The observations are carried out at the {\it Hobby-Eberly
Telescope} and at the {\it Wise Observatory}. The data cover 8 yr of
photometric monitoring of 11 quasars, and 2.5 yr of spectrophotometric
monitoring of 7 of these sources. Thus far we detected continuum
variations but no line variations. We find that the continua of the
high-luminosity quasars have smaller variability amplitudes and longer
variability timescales compared with low-luminosity AGNs.
\end{abstract}
\vglue -0.6cm

\vglue -0.6cm
\section{Introduction}

\vglue -0.2cm
While the origin of the continuum variability of Active Galactic
Nuclei (AGN) is still unclear, it is possible to use reverberation
mapping to study AGN emission-line gas. In such studies, the time lags
between the variations in the continuum flux and the emission-lines'
fluxes are used to estimate the Broad Line Region's (BLR) size and
map its geometry. Twenty low-luminosity AGN (Seyfert~1 galaxies) were
spectrophotometrically monitored and their BLR sizes were measured
using this method. Kaspi et~al. (2000, ApJ, 533, 631) measured
BLR sizes for 17 high-luminosity AGN (quasars) which increased
the luminosity range of objects with known BLR size by two orders
of magnitude. Kaspi et~al. used all those BLR size measurements to
determine for the first time the size--mass--luminosity relations
over a luminosity range of 5 orders of magnitude.

The above sample is luminosity and redshift limited because of
the small telescopes used. In order to increase the luminosity
beyond 10$^{46}$~erg\,s$^{-1}$ we need to monitor high-redshift,
high-luminosity quasars.
\vglue -0.2cm

\begin{figure}[t]
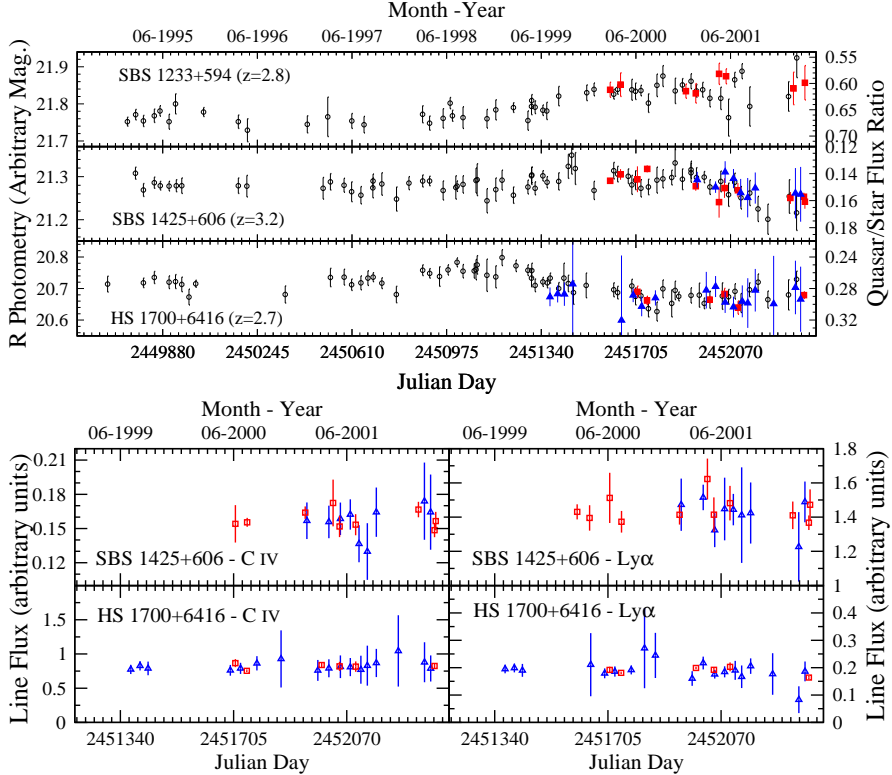

\centerline{
\psfig{figure=kaspi1.eps,width=11.75cm}
}
\centerline{
\psfig{figure=kaspi2.eps,width=11.75cm}
}
\vglue -1.5cm
\caption{{\it Top}: 
Continuum light curves of three high-luminosity quasars. Open
circles are data obtained from $R$-band photometry at the {\it Wise
Observatory} over 8 yr. The $\sim$\,0.1 magnitude variations are
apparent in all objects. Squares are data obtained from {\it HET}
spectra, and triangles are data obtained from {\it Wise} spectra over
the past 2.5 yr. These data are represented by the quasar/star flux
ratios which were shifted and stretched to fit the $R$-band photometry
light curves. For most objects the shapes of the light curves from
the two data sets are similar.
{\it Bottom}:
Light curves for C\,{\sc iv} and Ly$\alpha$ for two objects. Symbols
are as above. No line variations are detected in the data.
}
\label{fig1}
\vglue -0.4cm
\end{figure}

\section{Monitoring Program and Preliminary Results} 

\vglue -0.2cm
For the past 8 yr we have been monitoring photometrically a sample
of 11 high-redshift (2.1 {$\la$} {$z$} {$\la$} 3.2), high-luminosity
(10$^{45.6}$ {$\la$} {$\lambda L_{\lambda}(5100{\mbox{\AA}})$} {$\la$}
10$^{47}$ erg\,s$^{-1}$) quasars. The choice of the objects in this
sample is purely observational (limits on redshift, declination,
observed flux, etc.). The observations are carried out every month at
the {\it Wise Observatory} using broad-band filters ($B$ \& $R$); thus
we have good coverage of continuum variability. Most of the objects
show 10\% peak-to-peak continuum variations (Fig.~1 top). For the past
2.5 yr we have also monitored part of this sample, spectroscopically,
with the 9~m {\it Hobby-Eberly Telescope (HET)} and the {\it Wise
Observatory} telescope to check for the line (Ly$\alpha$ and C{\sc
iv}$\lambda 1549$) response to the continuum variations.

First results indicate that although continuum variations are detected
no line variations are yet seen (Fig. 1 bottom). Our typical upper
limit for the line variability is $\la 25$\%. This is not surprising
since these high-luminosity quasars are expected to have BLR sizes
of order 3 light years which correspond, in our frame, to about a
decade. In addition, significant continuum variation are needed to
produce a detectable line response. Comparison with low-luminosity
AGN shows that the continua of the high-luminosity quasars have
smaller variability amplitudes and longer variability timescales. We
expect that, when this long-term project is concluded, reverberation
measurements of the BLR size will cover the luminosity range of
10$^{41}$--10$^{47}$ ergs\,s$^{-1}$.

\end{document}